\documentclass[]{aa}
\usepackage{graphicx}
\usepackage{amssymb}
\begin{document}
   \title{Modeling $\beta$ Virginis using seismological data}

   \author{P. Eggenberger \and F. Carrier}
   \institute{Observatoire de Gen\`eve, 51 chemin des Maillettes,
             CH-1290 Sauverny, Switzerland}
       
   \offprints{P. Eggenberger\\
   \email{Patrick.Eggenberger@obs.unige.ch}}
  
  \date{Received ; accepted }
  \titlerunning{Modeling of $\beta$ Virginis}
  \abstract{This paper presents the modeling of the F9 V star \object{$\beta$ Virginis} based on seismological
  measurements. Using the Geneva evolution code including rotation and atomic diffusion, 
  we find that two distinct solutions reproduce all existing asteroseismic and non-asteroseismic 
  observational constraints well:
  a main-sequence model with a mass of $1.28 \pm 0.03$\,$M_{\odot}$ and an age $t=3.24 \pm 0.20$\,Gyr, 
  or a model in the post-main sequence phase of evolution with a lower mass of $1.21 \pm 0.02$\,$M_{\odot}$ 
  and an age $t=4.01 \pm 0.30$\,Gyr.
  The small spacings $\delta \nu_{02}$ and the ratio $r_{02}$ between small and
  large spacings are sensitive to the differences in the structure of the central layers between these two solutions
  and are also sensitive to the structural changes due to the rotational mixing.
  They can therefore be used to unambiguously determine the evolutionary state of \object{$\beta$ Vir} and 
  to study the effects of rotation on the inner structure of the star.
  Unfortunately, existing asteroseismic data do not enable such precise determination.
  We also show that the scatter in frequencies introduced by the rotational splittings can account for the larger dispersion
  of the observed large spacings for the non-radial modes than for the radial modes.
}
  \maketitle
%

\section{Introduction}

The solar five-minute oscillations have led to a wealth of information
about the internal structure of the Sun. 
These results stimulated various attempts to detect
a similar signal on other solar-like stars by photometric or equivalent
width measurements. In past years, the stabilized spectrographs developed for an extra-solar planet 
search have finally achieved the accuracy needed to make these detections. 
While solar-like oscillations have been detected for a handful of solar-type stars,
individual p-mode frequencies have only been identified for a few of these stars:
$\alpha$ Cen A (Bouchy \& Carrier \cite{bo02}; Bedding et al. \cite{be04}), $\alpha$ Cen B 
(Carrier \& Bourban \cite{ca03}; Kjeldsen et al. \cite{kj05}),
Procyon A (Marti\'c et al. \cite{ma04a}; Eggenberger et al. \cite{eg04a}),
$\eta$ Bootis (Kjeldsen et al. \cite{kj03}; Carrier et al. \cite{ca05a}),
and $\mu$ Arae (Bouchy et al. \cite{bo05}).
Individual frequencies have also been identified in the giant star $\xi$\,Hydrae (Frandsen et al. \cite{fr02}).

Based on these asteroseismic data, numerous theoretical analyses have been performed in order 
to determine precise global stellar parameters (see for instance Eggenberger et al. \cite{eg04b} for the
$\alpha$ Cen system) and to try to test the inclusion of overshooting or rotation in the models 
(see Di Mauro et al. \cite{di03} for $\eta$ Boo or Eggenberger et al. \cite{eg05} for Procyon).

Recently, Carrier et al. (\cite{ca05b}) detected solar-like oscillations on the F9 V star \object{$\beta$ Virginis} (HD 102870) and 
reported identification of several individual frequencies. 
Marti\'c et al. (\cite{ma04b}) also detected solar-like oscillations on \object{$\beta$ Virginis},
but no individual frequencies were identified.
The aim of the present paper is 
to investigate which additional constraints are brought by these seismological data. We will thus try to
determine the model of \object{$\beta$ Virginis} that best reproduces all the observational constraints using the Geneva
evolution code, which includes a comprehensive treatment of shellular rotation and atomic diffusion. 

The observational constraints available for \object{$\beta$ Virginis} are summarized in Sect.~2, 
while the computational method is described in Sect.~3. The results are presented in Sect.~4 and the effects of rotation are discussed
in Sect.~5. The conclusion is given in Sect.~6.

\section{Observational constraints}
\label{obs}

\subsection{Effective temperature and chemical composition}

From the different spectroscopic measurements reported in the literature 
for \object{$\beta$ Vir} (Edvardsson et al. \cite{ed93}; Buzzoni et al. \cite{bu01}; Gray et al. \cite{gr01}; 
Le Borgne et al. \cite{le03}; Taylor et al. \cite{ta03}; Morel \& Micela \cite{mo04}; Allende Prieto et al. \cite{al04};
Lambert et al. \cite{la04}), we adopt an effective temperature $T_{\mathrm{eff}}=6130 \pm 50$\,K and
a metallicity [Fe/H$]=0.14 \pm 0.05$.

\subsection{Luminosity}

Combining the magnitude $V = 3.597 \pm 0.004$ (Burki et al. \cite{bu05}),
the Hipparcos parallax $\Pi=91.74 \pm 0.77$\,mas, the solar absolute bolometric magnitude $M_{\mathrm{bol},\,\odot}=4.746$
(Lejeune et al. \cite{le98}) and the
bolometric corrections from Flower's
(\cite{flower}) $BC = -0.028 \pm 0.006$\,mag determined from the effective temperature, 
we find a luminosity $L=3.51 \pm 0.10$\,$L_{\odot}$.
 
\subsection{Rotational velocity}

We use the observed surface velocity of \object{$\beta$ Vir} 
to constrain the rotational velocity of our models.
Glebocki \& Stawikowski (\cite{gs00}) have determined a velocity $V \sin i \cong 4.3$\,km\,s$^{-1}$. 
Since the value of the angle $i$ is unknown, we assume that it is close to 90\,$\degr$. 
Thus our models of \object{$\beta$ Vir}
have to reproduce a surface velocity of about 4.3\,km\,s$^{-1}$. 

All non-asteroseismic observational constraints are listed in Table~\ref{tab:constraints}.

\subsection{Asteroseismic constraints}
\label{ascon}

Solar-like oscillations in \object{$\beta$ Vir} have been detected by Carrier et al. (\cite{ca05b}) 
with the \textsc{Coralie} echelle spectrograph.
Thirty-one oscillation frequencies have been identified in the power spectrum between 0.7 and 2.4\,mHz 
with amplitudes in the range 23 to 46\,cm~s$^{-1}$. 
By a least square fit of the asymptotic relation with the identified frequencies,
they determined a mean large and small separation of $\Delta \nu = 72.1$\,$\mu$Hz and 
$\delta \nu_{02}=6.3$\,$\mu$Hz, respectively. 
The precision of the frequency determination is estimated by conducting 
Monte Carlo simulations in which time series computed for various values of the damping time 
and re-excitation rates are analyzed. 
A velocity time series for detected oscillation modes is built by using the observational time
sampling and the observational noise on radial velocities. The amplitude spectrum of this   
series is then calculated and the differences between the simulated peaks, and the actual frequencies 
are determined.
The whole procedure is repeated 1000 times for each set of parameters
to ensure the stability of the results. These simulations lead to an estimation
of the error on individual frequency between 0.8 and 1.4\,$\mu$Hz, mainly depending
on the damping time. This result is very close to the frequency resolution
of the time series (1.14\,$\mu$Hz). In the following, we thus adopt an error on individual frequencies
estimated as the frequency resolution of the time series.

\begin{table}
\caption{Non-asteroseismic observational constraints for \object{$\beta$ Vir}.}

\begin{center}
\begin{tabular}{cc}
\hline
\hline
$\Pi$ [mas]& $91.74 \pm 0.77$  \\
$V$ [mag] & $3.597 \pm 0.004$  \\
$L/L_{\odot}$ & $3.51 \pm 0.10$ \\
$T_{\mathrm{eff}}$ [K]& $6130 \pm 50$ \\
$[$Fe/H$]$ &  $0.14 \pm 0.05$ \\
$V \sin i$ [km\,s$^{-1}$] & $\sim 4.3$ \\
\hline
\label{tab:constraints}
\end{tabular}
\end{center}
\end{table}

\section{Stellar models}
\label{comp}

\subsection{Code description}

\subsubsection{General input physics}

The stellar evolution code used for these computations is the Geneva code
that includes shellular rotation (Meynet \& Maeder \cite{mm00}).
We use the OPAL opacities, the NACRE nuclear reaction
rates (Angulo et al. \cite{an99}), and the standard mixing-length formalism for convection. 
We do not include an additional convective penetration from the convective core into the surrounding stable layers,
since the inclusion of shellular rotation results in an increase of the convective core (see Fig.~11 of Carrier et al. 
\cite{ca05a}).

\subsubsection{Rotation}

We follow the evolution of the rotation profile from the zero age main sequence on,
assuming initial solid body rotation. The braking law of Kawaler (\cite{ka88}) is used 
in order to reproduce the magnetic braking undergone by low
mass stars when arriving on the main sequence. Two parameters enter this braking law: the
saturation velocity $\Omega_{\rm{sat}}$ and the braking constant $K$. Following Bouvier et al.
(\cite{bo97}), $\Omega_{\rm{sat}}$ is fixed to 14\,$\Omega_{\odot}$ and the braking constant $K$
is calibrated on the Sun.

\subsubsection{Meridional circulation}

The velocity of the meridional circulation in the case of shellular
rotation was initially derived by Zahn (\cite{za92}).
The effects of the vertical $\mu$-gradient $\nabla_{\mu}$ and of the 
horizontal turbulence on meridional circulation were then taken into
account by Maeder \& Zahn (\cite{ma98}). They find
\begin{eqnarray}
U(r) & = & \frac{P}{\rho g C_{P} T [\nabla_{\rm ad}-\nabla + (\varphi/\delta)
  \nabla_{\mu}]} \nonumber\\
&  & \times  \left\{ \frac{L}{M}(E_{\Omega }+E_{\mu}) \right\}\,, 
\label{vmer}
\end{eqnarray}
where $P$ is the pressure and $C_P$ the specific heat. The $E_{\Omega}$ and $E_{\mu}$ terms
depend on the $\Omega$- and $\mu$-distributions, respectively, up to the third
order derivatives and on various thermodynamic quantities (see Maeder \& Zahn \cite{ma98}
for more details).

\subsubsection{Shear turbulence}

The diffusion by shear instabilities is expressed by a coefficient $D_{\rm shear}$:
\begin{eqnarray}
D_{\rm shear} & = & \frac{ 4(K + D_{\mathrm{h}})}
{\left[\frac{\varphi}{\delta} 
\nabla_{\mu}(1+\frac{K}{D_{\mathrm{h}}})+ (\nabla_{\mathrm{ad}}
-\nabla_{\mathrm{rad}}) \right] } \nonumber\\
&  & \times \frac{H_{\mathrm{p}}}{g \delta} \; 
\left [ \frac{\alpha}{4}\left( f\Omega{{\rm d}\ln \Omega \over {\rm d}\ln r} \right)^2
-(\nabla^{\prime}  -\nabla) \right] \, ,
\label{dshear}
\end{eqnarray}
where $f$ is a numerical factor equal to 0.8836, $K$ the thermal diffusivity,
and $(\nabla^{\prime}  -\nabla)$ expresses the difference between the internal 
nonadiabatic gradient and the local gradient (Maeder \& Meynet \cite{ma01}).

\subsubsection{Horizontal turbulence}

We used the new horizontal turbulence prescription of Maeder (2003)
that expresses the balance between the energy dissipated by the 
horizontal turbulence and the excess of energy present in the differential rotation
on an equipotential that can be dissipated in a dynamical time:
\begin{equation}
D_{\rm h} = A r \left( r \Omega(r) V [2V-\alpha U]\right)^{\frac{1}{3}} \,,
\label{Dhmaeder}
\end{equation}
with
\begin{equation}
A= \left( \frac{3}{400 n \pi} \right)^{\frac{1}{3}} \,,
\label{Amaeder}
\end{equation} 
where $U(r)$ is the vertical component of the meridional circulation velocity, $V(r)$ the
horizontal component, and $\alpha=\frac{1}{2} \frac{{\rm d} \ln r^2 \Omega}{{\rm d} \ln r}$.

\subsubsection{Transport of chemicals}

The vertical transport of chemicals
through the combined action of vertical advection and strong
horizontal diffusion 
can be described as a pure diffusive process (Chaboyer \& Zahn \cite{ch92}).
Indeed, the advective transport 
can be replaced by a diffusive term, with an effective
diffusion coefficient
\begin{equation}
D_{\rm eff} = \frac{|rU(r)|^2}{30D_{\rm h}}\,,
\label{Deff}
\end{equation}
where $D_{\rm h}$ is the coefficient of horizontal diffusion 
(see Eq.~\ref{Dhmaeder}).

Atomic diffusion driven by gravitational settling and thermal gradients
is included using the prescription by Paquette et al. (\cite{pa86}).
The evolution of an element concentration per unit mass $c_i$ then
follows:  
\begin{eqnarray}
\rho  \frac{{\rm d} c_i}{{\rm d} t} = \dot{c_i}
 + \frac{1}{r^2}\frac{\partial}{\partial r}\left[r^2\rho
   \left\{U_{\rm diff}c_i + (D_{\rm eff}+D_{\rm shear})\frac{\partial c_i}{\partial r}\right\}\right]
\label{transchem}
\end{eqnarray}
with $U_{\rm diff}$ the microscopic diffusion velocity 
and $\dot{c_i}$ the variations of chemical
composition due to nuclear reactions.

\subsection{Computational method}

The characteristics of a stellar model that includes the effects of rotation 
depend on six modeling parameters: the mass $M$ of the star, its age (hereafter $t$), 
the mixing-length parameter $\alpha \equiv l/H_{\mathrm{p}}$ for convection, 
the initial surface velocity $V_{\mathrm{i}}$,
and two parameters describing the initial chemical composition of the star. For these two parameters,
we choose the initial hydrogen abundance $X_{\mathrm{i}}$ and 
the initial ratio between the mass fraction of heavy elements and hydrogen 
$(Z/X)_{\mathrm{i}}$. 
This ratio is directly related to the abundance ratio [Fe/H] 
assuming that $\log(Z/X) \cong \mathrm{[Fe/H]} + \log(Z/X)_{\odot}$; 
we use the solar value $(Z/X)_{\odot}=0.0230$ given by Grevesse \& Sauval (\cite{gr98}).

Contrary to our previous seismic calibrations of 
binary stars like $\alpha$ Centauri and Procyon,
we decided to limit the parameter space of the present analysis 
by fixing the mixing-length parameter to its solar calibrated value ($\alpha_{\odot}=1.75$)
and the hydrogen abundance to $X=0.7$. 
This simply results from the limited number of non-asteroseismic 
observational constraints available for a star like \object{$\beta$ Virginis}, 
which does not belong to a binary system.
Although this arbitrarily limits our analysis to a 
small subset of possible solutions,
we think that it is the appropriate way to proceed 
for a first modeling of an isolated star in order to investigate
which constraints are brought by the observed p-mode frequencies.
A more thorough exploration of the parameter space, leading
to the independent determination of the mixing-length parameter and the hydrogen abundance,
will be possible when additional and more accurate seismic and non-asteroseismic observables 
(e.g. interferometric radius) are available for \object{$\beta$ Virginis}.

\begin{figure}[htb!]
 \resizebox{\hsize}{!}{\includegraphics{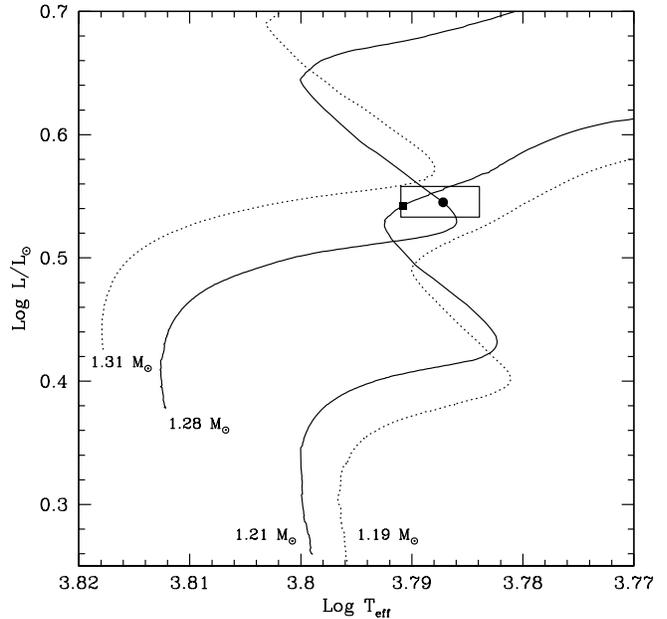}}
  \caption{Evolutionary tracks in the HR diagram for the two models of \object{$\beta$ Vir}.
The dot and the square indicate the location of the M1 and M2 models, respectively.
The error box in continuous line indicates the observational constraints of $L$ and $T_{\mathrm{eff}}$. The evolutionary
tracks of a 1.19\,$M_{\odot}$ and 1.31\,$M_{\odot}$ model are also plotted (dotted lines); this shows that models
with masses that are lower than $1.19$\,$M_{\odot}$ or larger than $1.31$\,$M_{\odot}$ are not able to reproduce the non-asteroseismic constraints.}
  \label{dhr}
\end{figure}

We then construct a grid of models with position in the HR diagram 
in agreement with the observational values of the luminosity and effective temperature (see Fig.~\ref{dhr}).
Note that the initial ratio between the mass fraction of heavy elements and hydrogen 
$(Z/X)_{\mathrm{i}}$ is directly constrained by the observed surface metallicity [Fe/H], while the initial velocity
$V_{\mathrm{i}}$ is directly constrained by the observed rotational velocity.
For each stellar model of this grid, low-$\ell$ p-mode frequencies are calculated using the Aarhus adiabatic 
pulsations package written by J. Christensen-Dalsgaard (\cite{cd97}). Following the observations, only modes $\ell \leq 2$
with frequencies between 0.7 and 2.4\,mHz are computed.
To determine the set of modeling parameters $(M,t,V_{\mathrm{i}},(Z/X)_{\mathrm{i}})$ that leads to the best agreement
with the observational constraints, two functionals are defined: $\chi^2_{\mathrm{clas}}$ and $\chi^2_{\mathrm{astero}}$.
The $\chi^2_{\mathrm{clas}}$ functional only uses the "classical" (i.e. non-asteroseismic) observables, and 
is defined as follows
\begin{eqnarray}
\label{eq11}
\chi^2_{\mathrm{clas}} \equiv \frac{1}{3} \sum_{i=1}^{3} \left( \frac{C_i^{\mathrm{theo}}-C_i^{\mathrm{obs}}}{\sigma C_i^{\mathrm{obs}}} \right)^2  \; ,
\end{eqnarray}
where the vector $\mathbf{C}$ contains the following observables for one star:  
\begin{eqnarray}
\nonumber
\mathbf{C} \equiv (L/L_{\odot},T_{\mathrm{eff}},[\mathrm{Fe/H}]) \; .   
\end{eqnarray} 
The vector $\mathbf{C}^{\mathrm{theo}}$ contains the theoretical values of these observables for the model to be tested, while 
the values of $\mathbf{C}^{\mathrm{obs}}$ are those
listed in Table~\ref{tab:constraints}. The vector $\mathbf{\sigma C}$ contains the errors on these observations.
Note that the observed rotational velocity is not included in this minimization, because of its large uncertainty resulting
from the unknown inclination angle $i$.
The $\chi^2_{\mathrm{astero}}$ functional compares individual theoretical frequencies to the observed ones.
The exact values of the computed p-mode frequencies depend on the details of the star's 
atmosphere, where the pulsation is non-adiabatic and where turbulent and radiative losses are significant.
The discrepancy between observed and theoretical solar p-mode frequencies calculated from standard 
models is indeed caused primarily by the approximations made in modeling the surface layers 
(Ulrich \& Rhodes \cite{ul83}; Christensen-Dalsgaard et al. \cite{cd88}; Guenther \cite{gu94};
Li et al. \cite{li02}).
The effects of poor modeling of the external layers
on computed p-mode frequencies can be taken into account by
using the discrepancies between observed and theoretical 
solar frequencies (see Guenther \& Brown \cite{gu04}). 
However, we do not know whether this estimation is reliable for stars with different 
surface gravity, chemical composition, and age.
Thus, we prefer to simply apply a constant shift
to correct theoretical frequencies for surface effects. 	  
We are well aware that this simple correction may seem doubtful, given
the strong frequency dependence of the solar near-surface effects,
but we think that this is the least biased 
procedure possible.
We therefore define the mean value of the differences between
the theoretical and observed frequencies:

\begin{eqnarray}
\nonumber
\langle D_{\nu}\rangle \equiv \frac{1}{N} \sum_{i=1}^N (\nu_i^{\mathrm{theo}}-\nu_i^{\mathrm{obs}}) \; ,
\end{eqnarray}
and then define the $\chi^2_{\mathrm{astero}}$ functional as
\begin{eqnarray}
\chi^2_{\mathrm{astero}} & \equiv & \frac{1}{N} \sum_{i=1}^{N} \left(
\frac{\nu_i^{\mathrm{theo}}-\nu_i^{\mathrm{obs}} - \langle D_{\nu}\rangle}{\sigma} \right)^2 \; ,
\label{eq2}
\end{eqnarray} 
where $\sigma=1.14$\,$\mu$Hz is the error on the observed frequencies, estimated as the frequency resolution
of the time series (see Sect.~\ref{ascon}), and $N$ is the number of observed frequencies.
Note that the adopted value of the error on the observed frequencies will of course change the values of
the computed $\chi^2_{\mathrm{astero}}$ (a larger value of $\sigma$ resulting in smaller values of 
$\chi^2_{\mathrm{astero}}$ and vice versa) but not the result of the minimization of the $\chi^2_{\mathrm{astero}}$ functional. 
Carrier et al. (\cite{ca05b}) report the identification of thirty-one individual frequencies. However,
four modes are identified at high frequencies and are thus far from the other modes, which makes
their identification less secure.
Moreover, these frequencies are found to be larger than the typical value of the acoustical cut-off
frequency of our models of \object{$\beta$ Vir} ($\nu_{\mathrm{ac}} \cong 2.1$\,mHz). 
As a result, these four frequencies are not
considered in our minimization, and $N$ is hence equal to 27 instead of 31.

Determination of the best set of parameters
is then based on the simultaneous minimization of the two functionals defined in Eqs.~\ref{eq11}
and \ref{eq2}. Finally, the global agreement between the stellar models and all classical and asteroseismic
observational constraints available for \object{$\beta$ Vir} is tested by defining the $\chi^2_{\mathrm{tot}}$
functional as the sum of $\chi^2_{\mathrm{clas}}$ and $\chi^2_{\mathrm{astero}}$:
\begin{eqnarray}
\chi^2_{\mathrm{tot}} \equiv \chi^2_{\mathrm{clas}} + \chi^2_{\mathrm{astero}} \; .
\label{eq3}
\end{eqnarray}

\section{Results}

Using the observational constraints listed in Sect.~\ref{obs} with the observed frequencies 
of Carrier et al. (\cite{ca05b}),
we perform the $\chi^2$ minimization described above. 

\begin{figure}[htb!]
 \resizebox{\hsize}{!}{\includegraphics{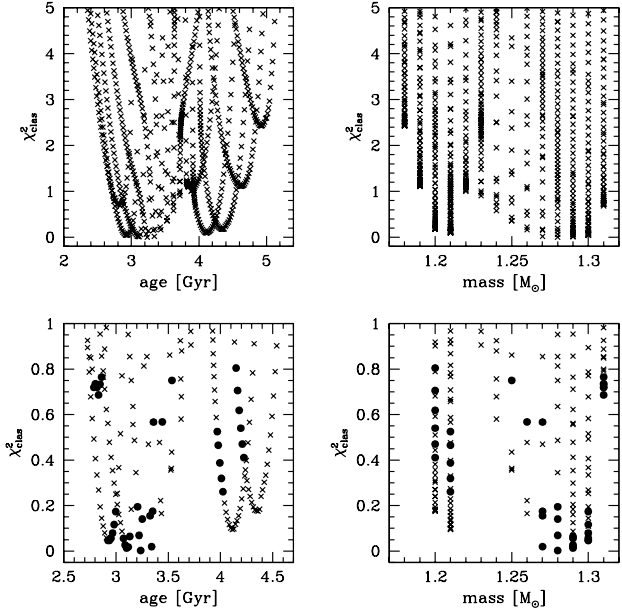}}
  \caption{{\bf Top:} $\chi^2_{\mathrm{clas}}$ of the models as a function of age and mass.
           {\bf Bottom:} Zoom of the above figures. Models with $\chi^2_{\mathrm{astero}} > 10$ are
	   still denoted by crosses, while models with $\chi^2_{\mathrm{astero}} \leq 10$ are
	   indicated by dots.}
  \label{kiclas}
\end{figure}

First, only the non-asteroseismic constraints are considered. Thus the $\chi^2_{\mathrm{clas}}$
minimization, which tests the consistency between 
observed and theoretical effective temperature, luminosity and metallicity, 
is applied. In this way, we find that two kind of solutions are able to reproduce the 
non-asteroseismic observational constraints of \object{$\beta$ Vir}:
one model with a mass of about 1.21\,$M_{\odot}$ and an age of about 4.1\,Gyr, and
a more massive model of 1.28\,$M_{\odot}$ with a smaller age of about 3.2\,Gyr. This is shown in
Fig.~\ref{kiclas}, which plots the values of $\chi^2_{\mathrm{clas}}$ for models matching the
observed surface metallicity of \object{$\beta$ Vir} as a function of age and mass. This figure
clearly shows one minimum near 1.21\,$M_{\odot}$ (corresponding to the minimum at 4.1\,Gyr) and another
minimum near 1.28\,$M_{\odot}$ (3.2\,Gyr).
To understand why two distinct solutions are found to fall within the observed constraints on 
effective temperature, luminosity and surface metallicity, we plot the evolutionary tracks in the
HR diagram corresponding to the 1.21 and 1.28\,$M_{\odot}$ models (see Fig.~\ref{dhr}).
Figure~\ref{dhr} shows that the two solutions correspond in fact to two different evolutionary stages:
the less massive model of 1.21\,$M_{\odot}$ has exhausted its hydrogen core and is therefore in the post-main
sequence phase of evolution, while the 1.28\,$M_{\odot}$ is still on the main sequence but is very
near hydrogen exhaustion. Since these two models have the same surface metallicity, the same surface velocity
($V \cong 4$\,km\,s$^{-1}$), and the same position in the HR diagram, non-asteroseismic observational
constraints do not enable us to determine the evolutionary
stage of \object{$\beta$ Vir}. Asteroseismic observations are needed to differentiate models with different
internal structure located in the same region of the HR diagram. Thus, we now examine which additional
constraints are brought by the asteroseismic measurements by performing the whole 
minimization described in Sect.~\ref{comp}.

\begin{figure}[htb!]
 \resizebox{\hsize}{!}{\includegraphics{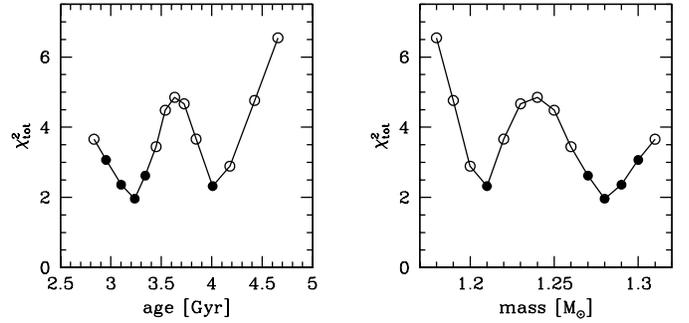}}
  \caption{Lowest values of $\chi^2_{\mathrm{tot}} \equiv \chi^2_{\mathrm{clas}} + \chi^2_{\mathrm{astero}}$ as a function of age and mass.
           Dots indicate models in accordance with the observed luminosity and effective temperature, while open circles
	   correspond to models that do not reproduce all non-asteroseismic observables.}
  \label{kitot}
\end{figure}

For this purpose, the $\chi^2_{\mathrm{astero}}$ values, which directly compare individual theoretical frequencies
to the observed ones, are computed. The bottom of Fig.~\ref{kiclas} shows which models are in reasonable agreement with
the observed p-mode frequencies (with $\chi^2_{\mathrm{astero}} \leq 10$) and which models do not reproduce the 
asteroseismic constraints ($\chi^2_{\mathrm{astero}} > 10$). We see that models still on the main-sequence, which
minimize $\chi^2_{\mathrm{clas}}$ for a mass and a corresponding age of about 1.28\,$M_{\odot}$ and 3.2\,Gyr respectively,
are found to also agree well with the seismological measurements, and hence to minimize
$\chi^2_{\mathrm{clas}}$ and $\chi^2_{\mathrm{astero}}$ at the same time. However, models in the post-main sequence phase of evolution,
which minimizes $\chi^2_{\mathrm{clas}}$ for a mass and a corresponding age of about 1.21\,$M_{\odot}$ and 4.1\,Gyr, have
values of $\chi^2_{\mathrm{astero}}$ that are higher than ten and therefore badly reproduce the asteroseismic
observables. Thus, post-main sequence models are not able to minimize the two functionals 
$\chi^2_{\mathrm{clas}}$ and $\chi^2_{\mathrm{astero}}$  at the same time as expected for a fully consistent model of \object{$\beta$ Vir}.
However, the bottom of Fig.~\ref{kiclas} shows that there are post-main sequence models that are compatible with
the observed frequencies and that also have small values of $\chi^2_{\mathrm{clas}}$, even if they do not minimize
$\chi^2_{\mathrm{clas}}$. This means that a post-main sequence model of \object{$\beta$ Vir} exists with position in the
HR diagram in agreement with the observed luminosity and effective temperature, which also
reproduces the observed $p$-mode frequencies well. This can be seen clearly by plotting the lowest values of the
$\chi^2_{\mathrm{tot}}$ functional, which tests the global agreement between the models and all observational
constraints, as a function of age and mass (see Fig.~\ref{kitot}). Figure~\ref{kitot} shows that there is a clear solution
near 1.28\,$M_{\odot}$ that minimizes all $\chi^2$ functionals at the same time. Figure~\ref{kitot}
also shows that the post-main sequence models near 1.21\,$M_{\odot}$ have low values of $\chi^2_{\mathrm{tot}}$ and are therefore
in good agreement with all observational constraints. At this stage of the analysis, we conclude that the 
additional asteroseismic measurements favor the solution
of a model still on the main sequence, but do not rule out the other solution of a less massive model in the post-main
sequence phase of evolution.       

Thus, two different solutions are determined:
a main-sequence model (denoted model M1 in the following) 
and a post-main sequence model (the M2 model).
For each model, the location in the HR diagram is shown in Fig.~\ref{dhr} and
the respective characteristics given in Table~\ref{tab:res}.
The confidence limits of each modeling parameter are estimated as the maximum/minimum values that 
fit the observational constraints when the other calibration parameters are fixed to their mean values.
Both models have a similar value of $\chi^2_{\mathrm{astero}}$, while the M2 model has a higher value of
$\chi^2_{\mathrm{clas}}$, which is mainly due to its high effective temperature (see Fig.\ref{dhr}).

We recall that these results have been obtained by fixing
the mixing-length parameter to its solar calibrated value ($\alpha_{\odot}=1.75$)
and the hydrogen abundance to $X=0.7$, since we think that current observational data do not
allow an independent determination of these quantities. We nevertheless investigated the
effect of a change to these values on the determined global parameters. Increasing the
mixing-length parameter $\alpha$ by 10\,\% for a given value of the hydrogen abundance
results in an increase of about 0.1 in the $\chi^2_{\mathrm{clas}}$ in order to
reach the same value of $\chi^2_{\mathrm{astero}}$; this also
decreases the mass by about 2\,\% and increases the age by about 10\,\%.
In the same way, increasing the hydrogen abundance by 10\,\% for a fixed value 
of the mixing-length parameter leads to an increase of about 0.7 in the $\chi^2_{\mathrm{clas}}$
in order to reach the same value of $\chi^2_{\mathrm{astero}}$; this 
increases the mass by about 12\,\% and 
decreases the age by about 6\,\%.
For a given mass, an increase in the hydrogen abundance can also be compensated for an increase 
in the mixing-length parameter and vice versa 
in order to reach the same location in the HR diagram (see Fig.~3 of Eggenberger et al. \cite{eg05}).

\begin{table}
\caption[]{Models for \object{$\beta$ Vir} including rotation and atomic diffusion. The upper part of the table gives the 
non-asteroseismic observational constraints used for the
calibration. The middle part of the table presents the modeling parameters with their confidence limits, while the bottom
part presents the global parameters of the star.}
\begin{center}
\label{tab:res}
\begin{tabular}{c|cc}
\hline
\hline
 & Model M1 & Model M2 \\ \hline
$L/L_{\odot}$ & \multicolumn{2}{c}{$3.51 \pm 0.10$} \\
$T_{\mathrm{eff}}$ [K]& \multicolumn{2}{c}{$6130 \pm 50$} \\
$[$Fe/H] & \multicolumn{2}{c}{$0.14 \pm 0.05$} \\
$V$ [km\,s$^{-1}$] & \multicolumn{2}{c}{$\sim 4.3$} \\
\hline
$M$ $[M_{\odot}]$ &  $1.28 \pm 0.03$ & $1.21 \pm 0.02$ \\
$t$ [Gyr] &  $3.24 \pm 0.20$ & $4.01 \pm 0.30$ \\
$V_{\mathrm{i}}$ [km\,s$^{-1}$] & $\sim 18 $& $\sim 18$\\
$(Z/X)_{\mathrm{i}}$ & $0.0340 \pm 0.0040$ & $0.0340 \pm 0.0040$   \\
\hline
$\chi^2_{\mathrm{clas}}$ & $0.002$ & $0.319$ \\
$\chi^2_{\mathrm{astero}}$ & $1.961$ & $2.003$ \\
$\chi^2_{\mathrm{tot}}$ & $1.963$ & $2.322$ \\
$L/L_{\odot}$ & $3.51$ & $3.49$ \\
$T_{\mathrm{eff}}$ [K]& $6126$ & $6177$ \\
$R/R_{\odot}$ & $1.666$ & $1.634$ \\
$[$Fe/H] & $0.14$ & $0.14$ \\
$(Z/X)$ & $0.0317$ & $0.0317$ \\
$V$ [km\,s$^{-1}$] & $4.2$ & $3.9$\\
$\Delta \nu$ [$\mu$Hz] & $72.1$  & $72.1$ \\
$\delta \nu_{02}$ [$\mu$Hz] & $5.4$ & $5.5$ \\
\hline
\end{tabular}
\end{center}
\end{table}

\begin{figure}[htb!]
 \resizebox{\hsize}{!}{\includegraphics{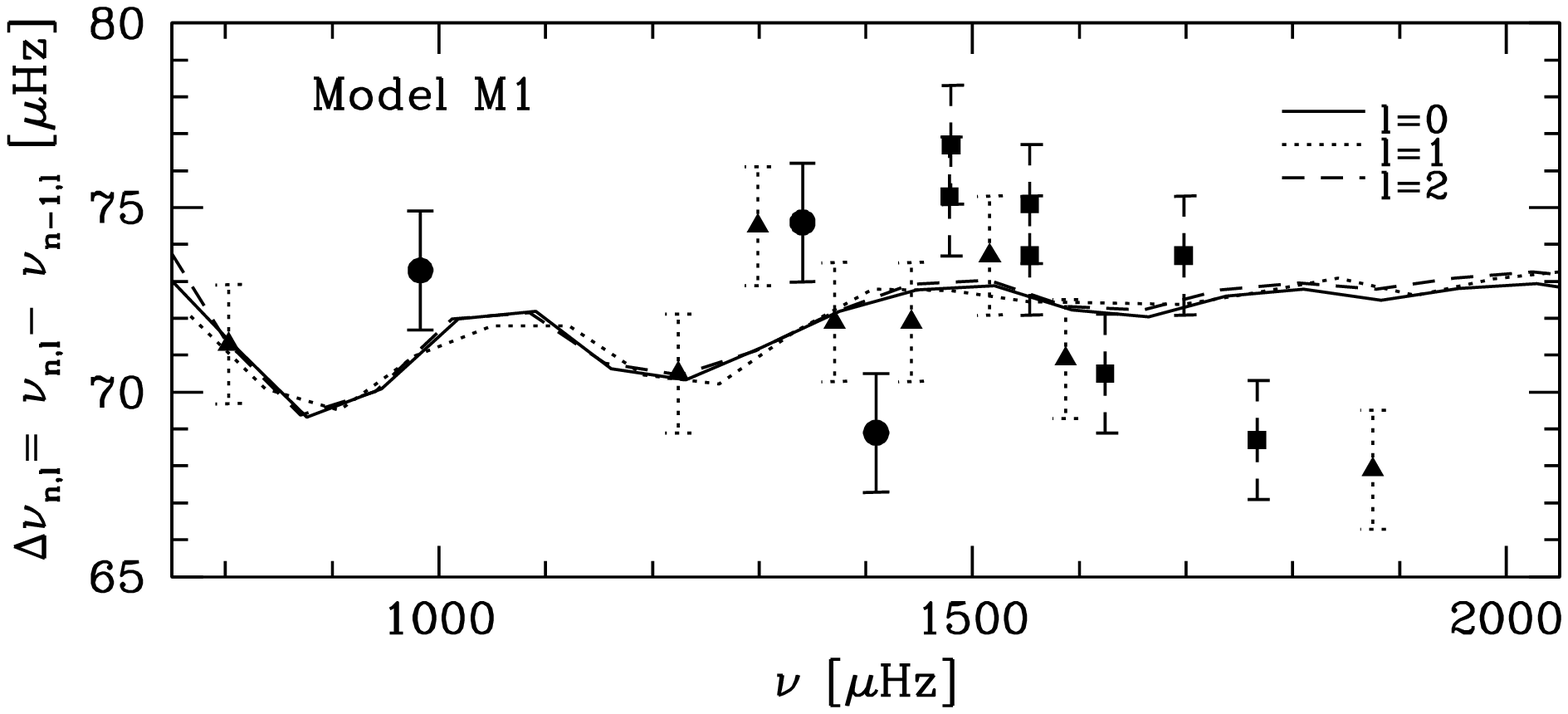}}
 \resizebox{\hsize}{!}{\includegraphics{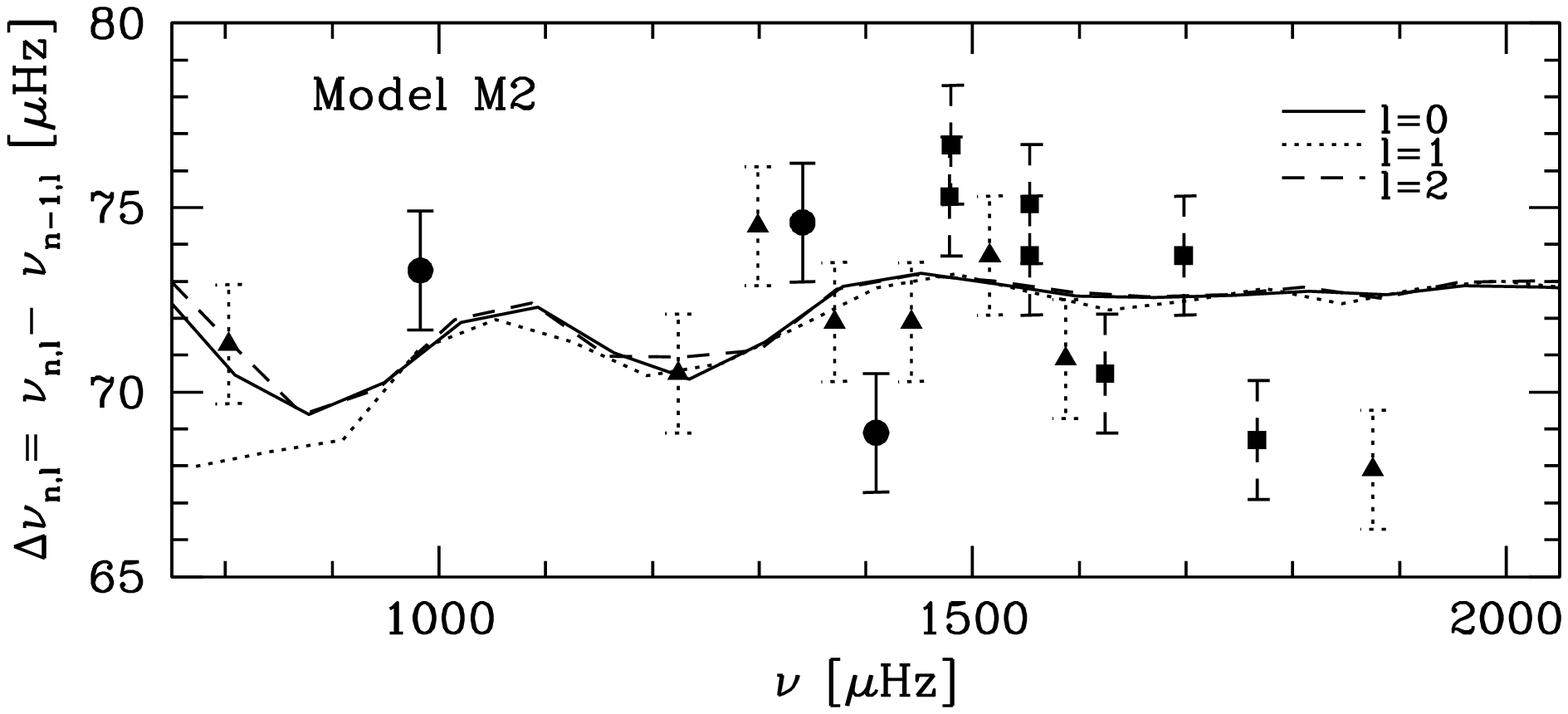}}
   \caption{Large spacings versus frequency for the M1 and M2 models.
  Dots indicate the observed values of the large spacing for radial modes, while triangles and squares correspond
  to the observed large spacings determined from $\ell=1$ and $\ell=2$ modes, respectively.}
  \label{gde}
\end{figure}

To study the asteroseismic features of the models, we first plot the  
variation of the large spacing ($\Delta \nu_{n,\ell} \equiv \nu_{n,\ell} - \nu_{n-1,\ell}$) with frequency; the comparison between observed and theoretical large 
spacings for the M1 and M2 model is given in Fig.~\ref{gde}. Table~\ref{tab:res} and Fig.~\ref{gde} show that the mean large spacing of both models are
in good agreement with the observed value of 72.1\,$\mu$Hz. Note that the radius of the M2 model is less than the one
of the M1 model in order to compensate for the decrease in mass and to reproduce the observed value of the mean large spacing.
The variations in the theoretical large spacings with frequency are also very similar for both models.
The only difference is that the M2 model exhibits
slightly lower values of the $\ell=1$ large spacings at low frequency than does the M1 model.
As a result, the M1 model is able to reproduce the observed $\ell=1$ large spacing at 800\,$\mu$Hz,
whereas the theoretical value of the M2 model is lower than the observed one.
This is due to the fact that the M2 model does not reproduce the observed $\ell=1$ mode
at 732\,$\mu$Hz well (see Fig.~\ref{diaech_M2}).
Finally, we note that the dispersion of the observed large spacings around the theoretical curves
is greater than expected taking an uncertainty of 1.14\,$\mu$Hz, the frequency resolution of
the time series, on the frequency determination.

\begin{figure}[htb!]
 \resizebox{\hsize}{!}{\includegraphics{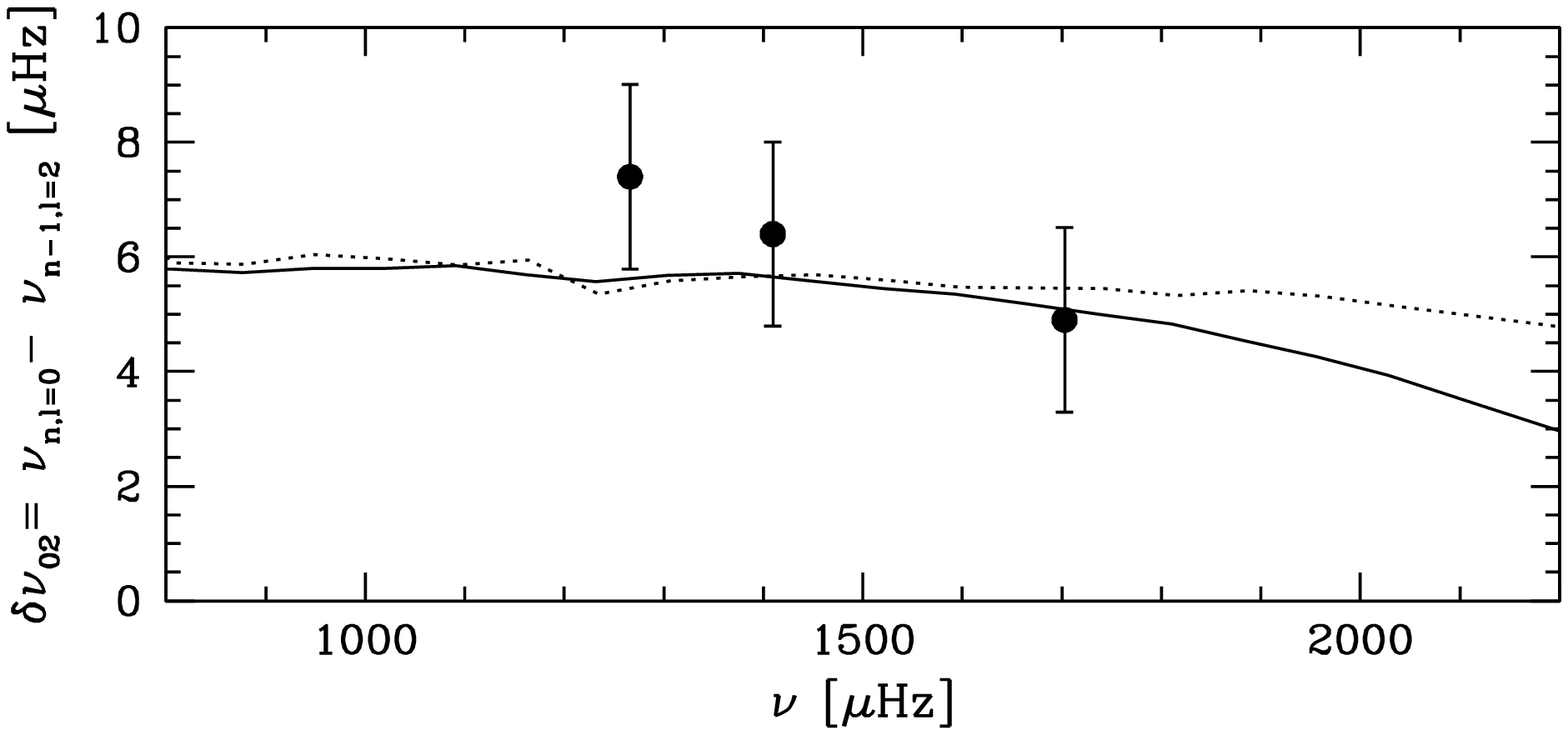}}
 \resizebox{\hsize}{!}{\includegraphics{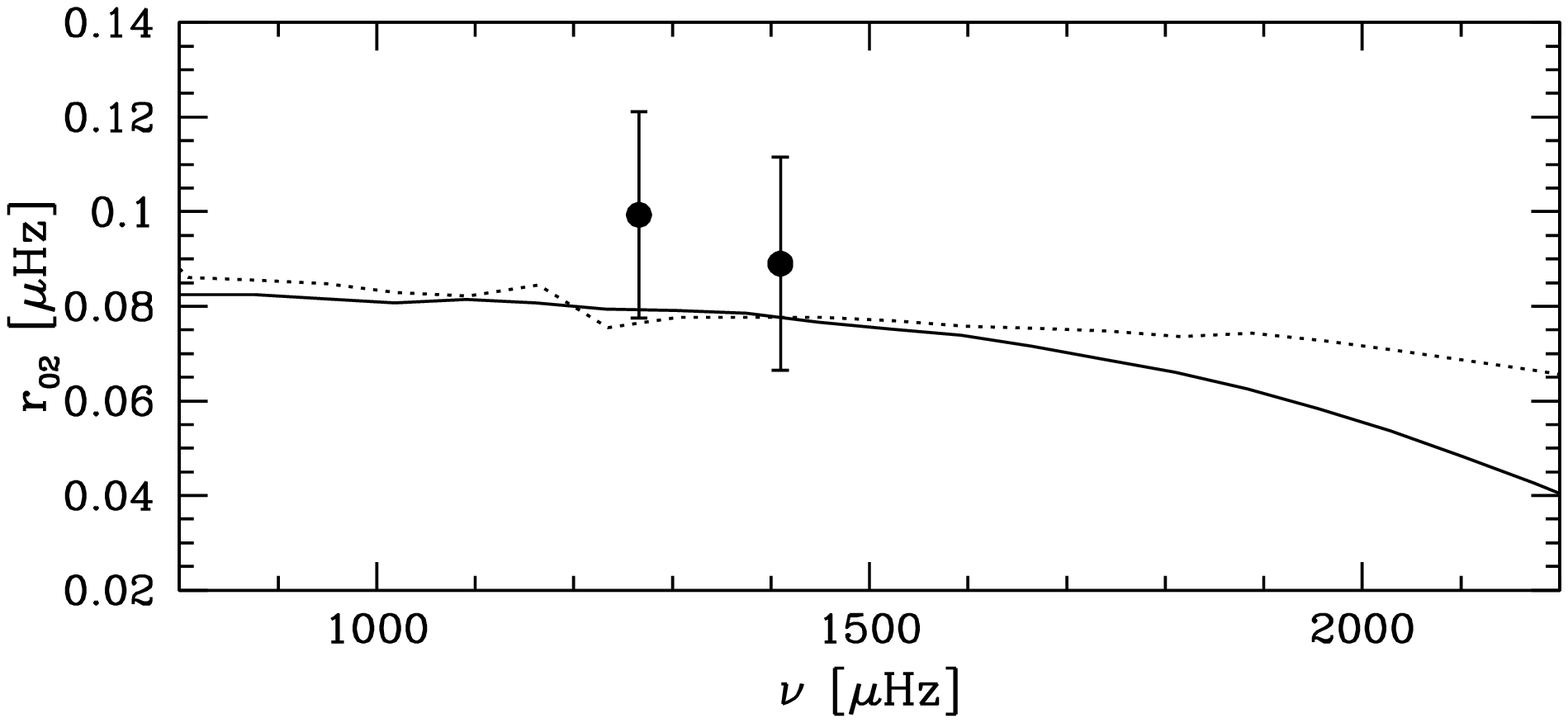}}
   \caption{Small spacing $\delta \nu_{02}$ and ratio of small to large separation 
            $r_{02} \equiv \delta \nu_{02}/\Delta \nu_{n,\ell=1}$ as a function of frequency.
	    The continuous line corresponds to the M1 model, while the dotted line
	    corresponds to the M2 model. Dots indicate the observed values.}
  \label{pte}
\end{figure} 

The small spacing $\delta \nu_{02} \equiv \nu_{n,\ell=0}- \nu_{n-1,\ell=2}$ is principally 
sensitive to the conditions in the central regions of the star. However, it also retains some sensitivity
to the near-surface structure. Roxburgh \& Vorontsov (\cite{ro03}) introduces the use of a new asteroseismic
diagnostic: the ratio of small
to large spacings $r_{02} \equiv \delta \nu_{02}/ \Delta \nu_{n,\ell=1}$. This ratio 
constitutes a better diagnostic of the central parts of a star than the small spacing,
since it is essentially independent of the structure of the outer layers, and is determined solely
by the interior structure (Roxburgh \& Vorontsov \cite{ro03}; Ot\'{\i} Floranes et al. \cite{ot04}).  

The variation in the small spacing $\delta \nu_{02}$ and the ratio $r_{02}$ with frequency for the M1 and
M2 models is given in Fig.~\ref{pte}. The large spacing corresponding to the observed small spacing near 1700\,$\mu$Hz has not
been observed; thus, only two observational values of the $r_{02}$ ratio are available.
We see that the theoretical mean small spacing of both models (see Table~\ref{tab:res}) is smaller than the observed value
of $6.3 \pm 1.4$\,$\mu$Hz.
Note that the theoretical mean small spacing has been determined by 
only considering the three theoretical values corresponding to the observed small spacings, in order to correctly
compare stellar models to observations. Figure~\ref{pte} shows that the difference between the theoretical and observed
mean small spacings is mainly due to the large observational value of $\delta \nu_{02}$ near 1270\,$\mu$Hz,     
which models fail to reproduce. This is also true for the $r_{02}$ ratio. 
The variation in $\delta \nu_{02}$ and $r_{02}$ with frequency is very similar
for both models at low frequency, but is found to be quite different at higher frequency.
Indeed, Fig.~\ref{pte} shows that the M1 and M2 models are characterized by approximately 
the same values of $\delta \nu_{02}$ and $r_{02}$ at frequencies lower than about
1600\,$\mu$Hz, while at higher frequencies, the M1 model exhibits a significantly 
stronger decrease in $\delta \nu_{02}$ and $r_{02}$ than does the M2 model.   
This is directly related to the evolutionary stage of the two models. Indeed, the post-main-sequence M2 model 
is more centrally condensed and therefore exhibits higher central values of pressure and density
than the younger M1 model, which is approaching hydrogen exhaustion but is still on the main-sequence.
Consequently, the sound speed and hence the Lamb frequency
of the two models differ in the central parts of the star.
The interior reflection of the $\ell=2$ modes are then different 
for the two models, which results in a stronger decrease in $\delta \nu_{02}$ and $r_{02}$ for the M1 model than for the M2 model.
Unfortunately, only one value for $\delta \nu_{02}$ and none for $r_{02}$ have been observed
for a frequency higher than 1600\,$\mu$Hz. The M1 model agrees slightly better with
this value of the small spacing near 1700\,$\mu$Hz than does the M2 model, but the difference is 
clearly not significant.  
  
\begin{figure}[htb!]
 \resizebox{\hsize}{!}{\includegraphics{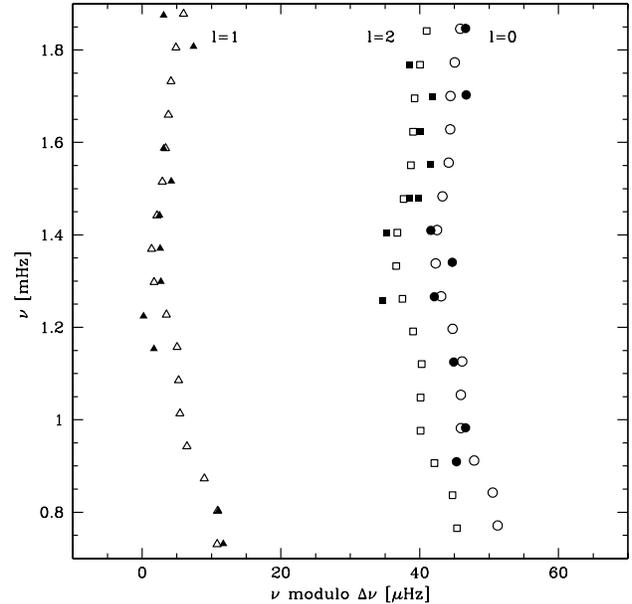}}
  \caption{Echelle diagram for the M1 model. 
  Open symbols refer to the theoretical frequencies, and filled symbols to the observed ones. 
  Circles are used for $\ell=0$ modes, triangles for $\ell=1$ modes, and squares for $\ell=2$ modes.
  The two $\ell=2$ modes near 1480\,$\mu$Hz and separated by 1.4\,$\mu$Hz
  could be due to rotational splitting.}
  \label{diaech_M1}
\end{figure}

\begin{figure}[htb!]
 \resizebox{\hsize}{!}{\includegraphics{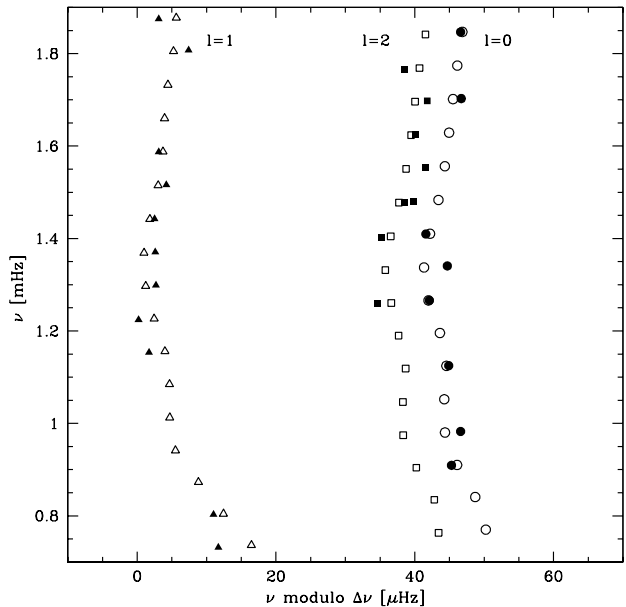}}
  \caption{Echelle diagram for the M2 model. 
  Open symbols refer to the theoretical frequencies, and filled symbols to the observed ones.  
  Circles are used for $\ell=0$ modes, triangles for $\ell=1$ modes, and squares for $\ell=2$ modes.}
  \label{diaech_M2}
\end{figure}

Finally, theoretical and observed p-mode frequencies are compared 
by plotting the echelle diagram of the M1 and M2 models (Figs.~\ref{diaech_M1} and \ref{diaech_M2}). 
In these figures the systematic difference $\langle D_{\nu}\rangle$ between theoretical and observed frequencies
has been taken into account ($\langle D_{\nu}\rangle = 23.8$\,$\mu$Hz and $21.1$\,$\mu$Hz 
for the M1 and M2 model, respectively). This shift of about 20\,$\mu$Hz between theoretical and observed
frequencies is substantially larger than the shift required in the solar case. We recall that our models have been computed 
with the standard mixing-length formalism for convection and that we assume the oscillations to be adiabatic. 
It is beyond the scope of the present paper to investigate the influence of these simple assumptions on the
solutions, but we note that they may explain a shift of about 20\,$\mu$Hz between 
computed and observed frequencies for a star that is more massive and more evolved than the Sun.

\section{Effects of rotation}

\subsection{Small spacings $\delta \nu_{02}$ and ratio $r_{02}$}

As already mentioned, the theoretical mean small spacings of the M1 and M2 models
are smaller than the observed ones. The theoretical values are, however, perfectly compatible with
the observations given the large uncertainty on the observed mean large spacing. Moreover, only
three values of $\delta \nu_{02}$ (and two of $r_{02}$) have been observed. It is nevertheless 
interesting to investigate which physical process can increase the theoretical value of the mean small
spacing in order to improve the agreement between the models and the seismological data.	    

Because the small spacing is principally 
sensitive to the structure of the core, its mean value decreases during the evolution of the star on the main sequence.
Thus, younger models than the M1 model, which is very near hydrogen exhaustion, exhibit higher values for the mean 
small spacing. Although these younger models are in better agreement with the observed values of $\delta \nu_{02}$, they fail
to reproduce the other seismological data. Indeed, the mean large spacing also decreases during the evolution 
of the star on the main sequence due to the increase of its radius. Consequently, the mean large spacing of 
the younger models is larger than the observed one; these models are therefore unable to reproduce the observed individual frequencies.

\begin{figure}[htb!]
 \resizebox{\hsize}{!}{\includegraphics{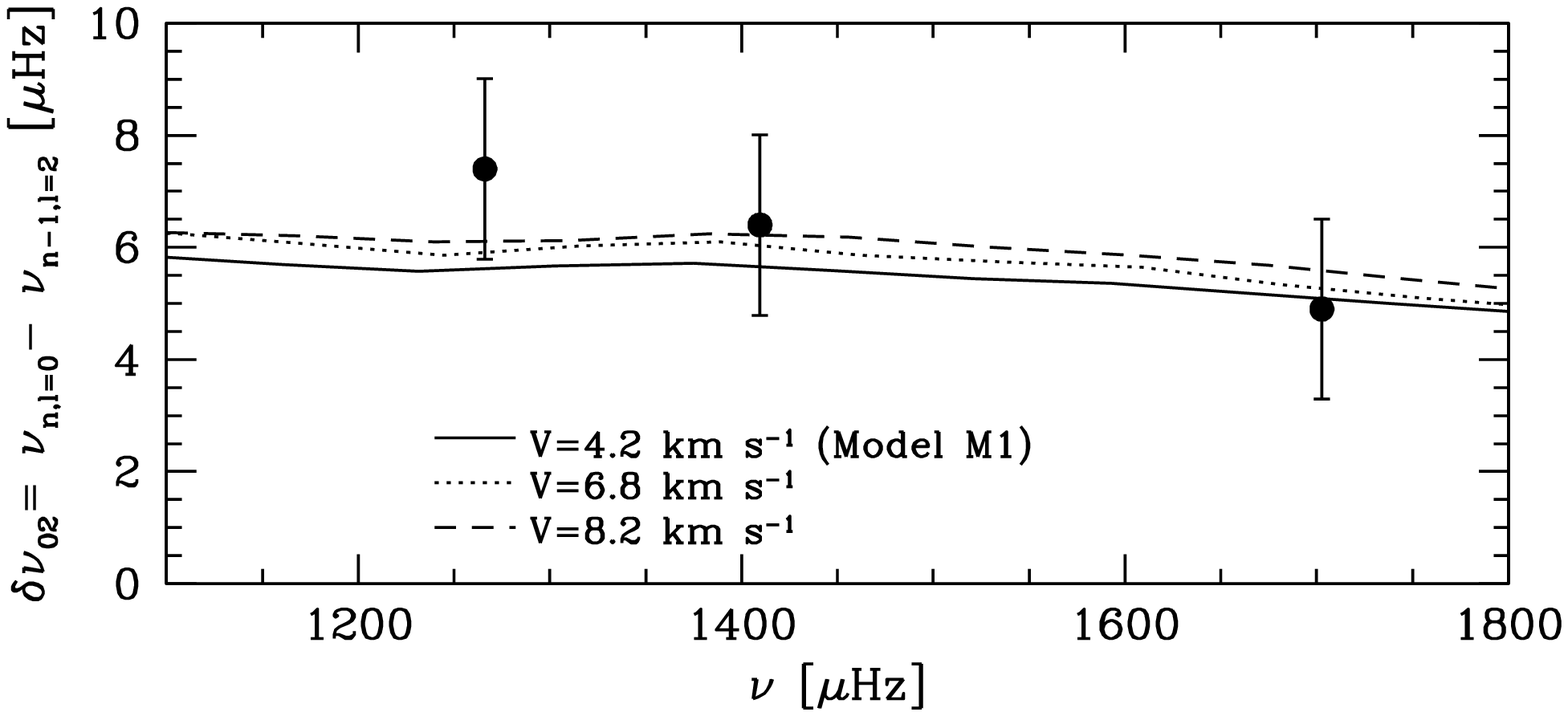}}
 \resizebox{\hsize}{!}{\includegraphics{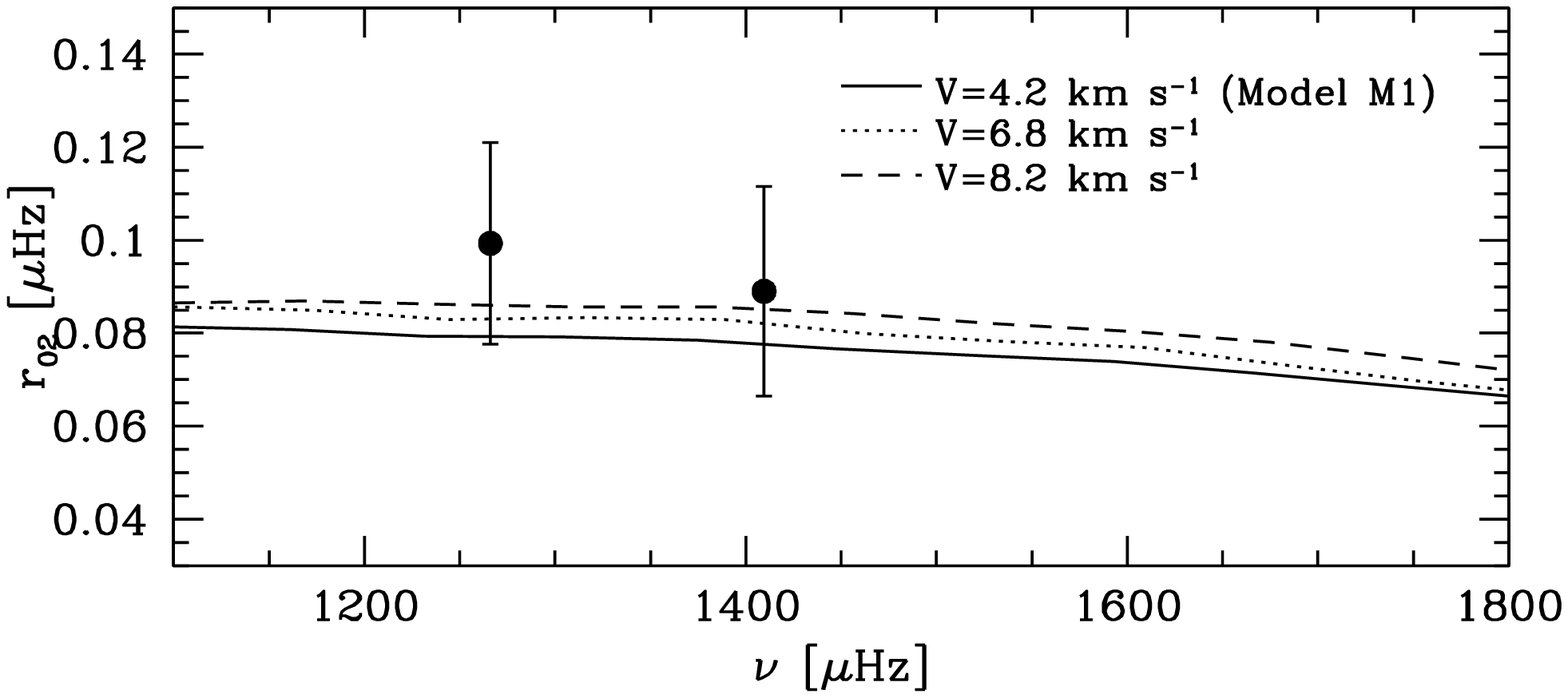}}
   \caption{Small spacing $\delta \nu_{02}$ and ratio of small to large separation 
            $r_{02} \equiv \delta \nu_{02}/\Delta \nu_{n,\ell=1}$ as a function of frequency for
	    models with the same value of the mean large spacing ($72.1$\,$\mu$Hz) but different surface
	    velocities $V$. Apart from the initial value of the rotational velocity, 
	    the models have been computed with the same initial parameters as the M1 model.  
            Dots indicate the observed values.}
  \label{pt_rot}
\end{figure} 

It is thus necessary to find a process that is able to increase the mean small spacing without changing the mean large
spacing. By changing the conditions in the central regions of the star, the rotational mixing is able to increase the values
of $\delta \nu_{02}$ and $r_{02}$ for models with the same mean large spacing. To investigate the effects of rotational 
mixing on $\delta \nu_{02}$ and $r_{02}$, we compute models with exactly the same initial parameters as the M1 model, but
different initial velocities. The M1 model is computed with an initial velocity of 18\,km\,s$^{-1}$ in order to reproduce
the observed surface velocity of about 4\,km\,s$^{-1}$. We calculate two other models with higher initial rotation velocities of 50 and
100\,km\,s$^{-1}$. Both models are evolved until their mean large spacing equals the observed value of 72.1\,$\mu$Hz.
The surface velocity $V$ decreases during the evolution on the main sequence as a result of the magnetic
braking that undergoes the low mass stars. When the observed value of the large spacing is reached, 
these velocities are equal to 6.8 and 8.2\,km\,s$^{-1}$ for the model with an initial velocity of 50 and 100\,km\,s$^{-1}$, respectively.

In this way, we obtain three models computed with the same input physics and initial parameters except for the initial velocity, and they
are in good global agreement with the asteroseismic measurements. The models with a higher initial velocity are of
course in poorer agreement with the classical constraints than the M1 model, since an increase in the rotational velocity results in an
increase in the effective temperature and luminosity. 
All models are, however, compatible with the adopted non-asteroseismic observational constraints.
The variation in the small spacing $\delta \nu_{02}$ and the ratio $r_{02}$ with frequency for the three models is given in Fig.~\ref{pt_rot}.
This figure shows that an increase in the rotational velocity increases $\delta \nu_{02}$ and $r_{02}$. Indeed, the mean small spacing
of the M1 model is 5.4\,$\mu$Hz, while it is equal to 5.7 and 6.0\,$\mu$Hz for the model with a surface velocity of 6.8 and 8.2\,km\,s$^{-1}$,
respectively. This illustrates the fact that the mixing in the central layers, hence the values of $\delta \nu_{02}$ and $r_{02}$, 
increase with the rotational velocity.

\subsection{Surface abundances}

\begin{figure}[htb!]
 \resizebox{\hsize}{!}{\includegraphics{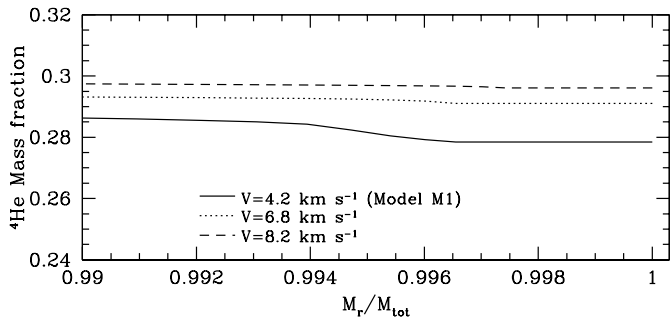}}
  \caption{Helium abundance profile in the external layers of the star 
  for models with the same value of the mean large spacing ($72.1$\,$\mu$Hz) but different surface
  velocities $V$.}
  \label{aby}
\end{figure}

In addition to changing the structure of the stellar core, rotation also influences the chemical abundances at the surface of the star.
Indeed, rotation-induced mixing counteracts the effects of atomic
diffusion in the external layers of the star. Since the efficiency of this mixing increases with the rotational velocity, the decrease of
the helium and heavy elements abundances due to atomic diffusion is found to be smaller for larger rotational velocities. 
This is illustrated in Fig.~\ref{aby}, which shows the helium profile in the external
layers of the star for the three rotating models with the same mean large spacing of 72.1\,$\mu$Hz. We see that the helium surface abundance
is larger when the surface velocity is higher. Although the effects of rotation on the surface abundances are relatively small for stars with masses
lower than 1.3\,$M_{\odot}$ like \object{$\beta$ Vir}, we note that they are particularly important
for more massive stars with shallower convective envelope
like Procyon or $\eta$ Bootis. 
Indeed, in stars that are more massive than about 1.4\,$M_{\odot}$, 
the rotation induced mixing prevents the helium and the heavy elements from being 
drained out of the convective envelope (see Fig.~10 of Carrier et al. \cite{ca05a} for $\eta$ Bootis).

\subsection{Dispersion of the observed large spacings}

\begin{figure}[htb!]
 \resizebox{\hsize}{!}{\includegraphics{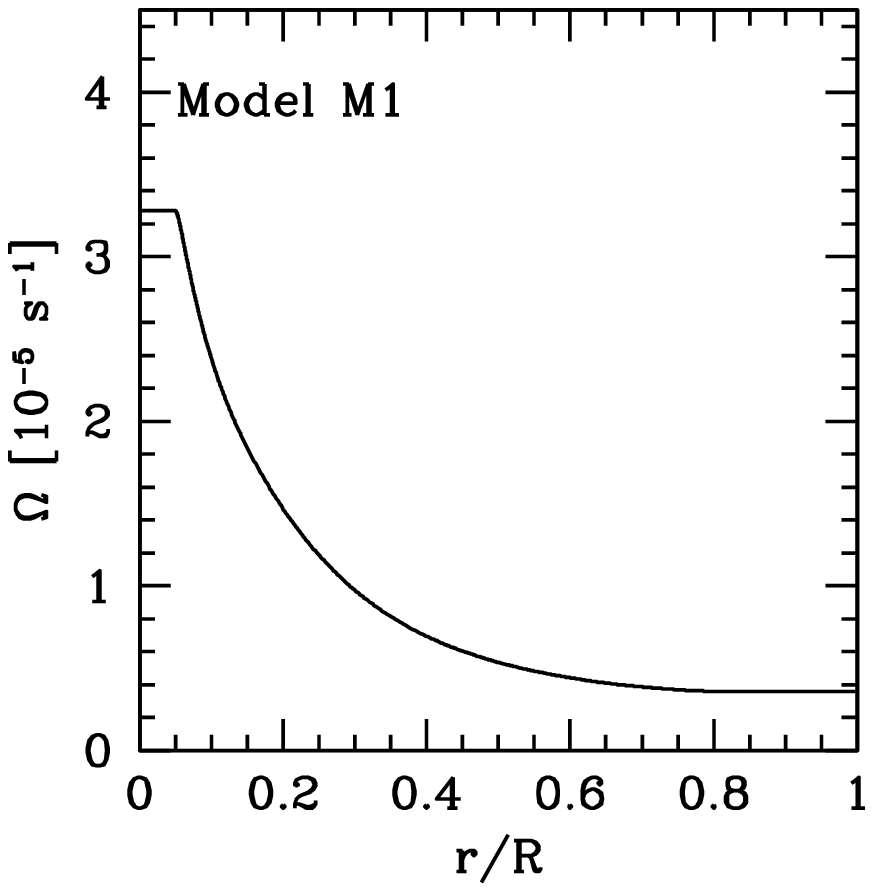} \includegraphics{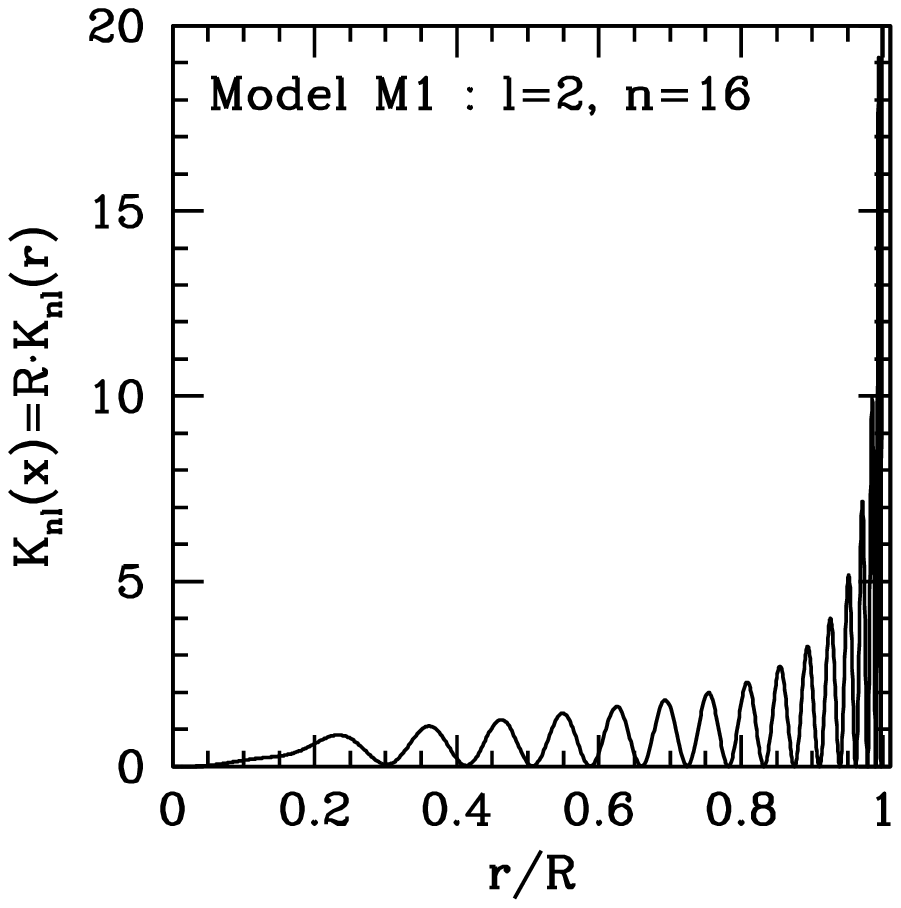}}
  \caption{Rotation profile of the M1 model and the rotational kernel corresponding to the $\ell=2$ mode
           with $n=16$.}
  \label{omega}
\end{figure}

\begin{figure}[htb!]
 \resizebox{\hsize}{!}{\includegraphics{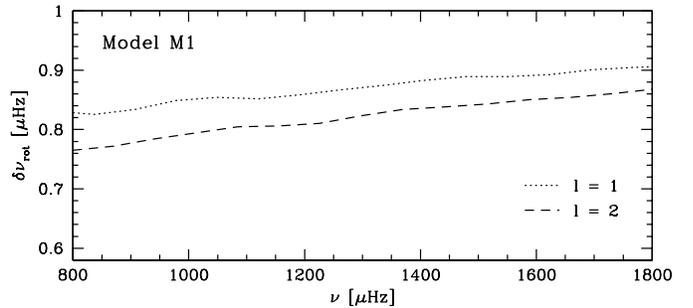}}
  \caption{Rotational splittings ($\delta \nu_{{\rm rot}}(n,\ell) \equiv \delta \omega_{n,\ell,m=1}/2\pi$) for the M1
  model. The dotted and the dashed lines correspond to the splittings for modes with $\ell=1$ and $\ell=2$, respectively.}
  \label{split}
\end{figure}

As mentioned above, the observed large spacings show a significant 
dispersion around the theoretical curves. Carrier et al. (\cite{ca05b}) suggest that this 
scatter can be partly explained by rotational splitting, since the dispersion for $\ell=2$ and
$\ell=1$ modes are found to be greater than for radial modes. To check this, we compute the theoretical rotational
splittings for the $\ell=2$ and $\ell=1$ modes of the M1 model. These frequency splittings are calculated using 
\begin{eqnarray}
  \delta \omega_{n\ell m}=m\beta_{n\ell} \int_{0}^{R}K_{n\ell}(r) \Omega(r)dr  \, ,
\label{rot}
\end{eqnarray}
where the rotational kernel $K_{n\ell}$ and the coefficients $\beta_{n\ell}$ are defined by Eqs. 
(8.42) and (8.43) from Christensen-Dalsgaard (\cite{cd03}). The rotation profile of the M1 model 
and an example of the rotational kernel computed for the $\ell=2$ mode with $n=16$ are
shown in Fig.~\ref{omega}. 
Using Eq. (\ref{rot}), the rotational splittings $\delta \nu_{{\rm rot}}$ expressed in terms of cyclic frequencies
($\delta \nu_{{\rm rot}}(n,\ell) \equiv \delta \omega_{n,\ell,m=1}/2\pi$) are calculated for the $\ell=1$ and $\ell=2$
modes; they are shown in Fig.~\ref{split}.
 
\begin{figure}[htb!]
 \resizebox{\hsize}{!}{\includegraphics{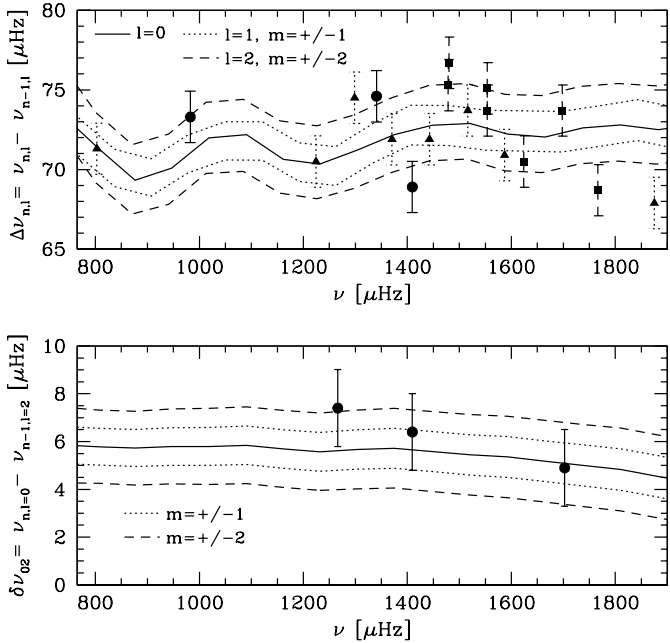}}
  \caption{Dispersion of the large and small spacings expected from rotational splittings for the M1 model. 
  The observed large spacings for the $\ell=1$ modes 
  are expected to lie between the maximum/minimum values delimited by the $\ell=1$, $m=\pm 1$ modes.
  In the same way, the observed large spacings for the $\ell=2$ modes and the observed small spacings 
$\delta \nu_{02}$ are expected to lie between the maximum/minimum values delimited by the $\ell=2$, $m=\pm 2$ modes.} 
  \label{gdpt_rot}
\end{figure}

The theoretical rotational splittings lie between 0.8 and 0.9\,$\mu$Hz for $\ell=1$ modes, and between 0.75 and 0.85\,$\mu$Hz for
$\ell=2$ modes. We notice that these values are higher than the value of about 0.6\,$\mu$Hz deduced by assuming a uniform rotation velocity
equal to the angular velocity at the surface. This reflects the increase of the angular velocity $\Omega$ when the distance
to the center of the star decreases (see Fig.~\ref{omega}). Due to the rotational splittings, 
the frequencies of the $\ell=1$ modes can be shifted by $\pm \delta \nu_{{\rm rot}}(\ell=1)$ ($\sim 0.85$\,$\mu$Hz),
and the frequencies of the $\ell=2$ modes can be shifted by $\pm 2\delta \nu_{{\rm rot}}(\ell=2)$ ($\sim 1.6$\,$\mu$Hz).    
This introduces a scatter of the frequencies which can explain the high dispersion of the observed large spacings.
Figure~\ref{gdpt_rot} shows the theoretical dispersion of the large spacings expected from rotational splittings; the values for the
$\ell=1$ modes are delimited by the modes with $m=\pm 1$, while the values for the $\ell=2$ modes
are delimited by the modes with $m=\pm 2$. 

We see that the scatter of frequencies resulting from the rotational splittings can account for the high dispersion
of the observed large spacings and especially for the fact that this dispersion is greater for $\ell=2$ and
$\ell=1$ modes than for radial modes.
Indeed, the observed large spacings for the $\ell=1$ and $\ell=2$ modes are found to lie within the
predicted interval in $\Delta \nu$. There is only one exception: the $\ell=1$ value
near 1880\,$\mu$Hz, which is too small to be correctly reproduced by the models. We also notice that the two values of 
the $\ell=0$ large spacing near 1400\,$\mu$Hz are not in good agreement with the theoretical predictions. 
This cannot, of course, been explained by rotational splittings, but can either be due to the scatter resulting from the 
finite lifetimes of the modes (which, contrary to the rotational splittings, affects all the modes including the radial ones)
or/and to the misidentification of the observed frequency near 1340\,$\mu$Hz. Indeed, the echelle diagrams show that the models fail to reproduce
the $\ell=0$ frequency near 1340\,$\mu$Hz (see Figs.~\ref{diaech_M1} and \ref{diaech_M2}). As a result, the two observed large spacings for $\ell=0$ modes near 1400\,$\mu$Hz 
do not agree well with the theoretical ones. 
It is interesting to note that the extracted frequency corresponding to the $\ell=0$ frequency at 1340.7\,$\mu$Hz
is 1329.1\,$\mu$Hz (see Table~2 in Carrier et al. \cite{ca05b}). Thus, this frequency has been shifted by $+11.57$\,$\mu$Hz by Carrier et al. 
and has been identified as an $\ell=0$ mode. Comparison with our models shows that the extracted frequency of 1329.1\,$\mu$Hz in fact better
agrees with the theoretical predictions than does the shifted frequency of 1340.7\,$\mu$Hz.
Thus, we think that this frequency, which was originally identified as a radial mode at 1340.7\,$\mu$Hz
by Carrier et al., is in fact an $\ell=2$ mode at 1329.1\,$\mu$Hz. 
The observed small spacings are also perfectly compatible with the theoretical predictions, when 
the uncertainties due to the rotational splittings of the $\ell=2$ modes are taken into account (see bottom of Fig.~\ref{gdpt_rot}).

Figure~\ref{omega} shows that models that include shellular rotation predict an increase in the angular velocity when the distance
to the center decreases in contrast to the near uniformity of the solar rotation profile. Indeed, helioseismological results indicate that
the angular velocity $\Omega(r)$ is constant as a function of the radius $r$
between about 20\,\% and 70\,\% of the total solar radius (Brown et al. \cite{br89}; Kosovichev et al. \cite{ko97};
Couvidat et al. \cite{co03}). Since shellular rotation alone produces an insufficient internal coupling to ensure
solid body rotation (Pinsonneault et al. \cite{pi89}; Chaboyer et al. \cite{ch95}; Eggenberger et al. \cite{eg05b}),
this suggests that another effect intervenes.
Two mechanisms have been proposed as reproducing the solar rotation profile: magnetic fields (Mestel \& Weiss \cite{me87};
Charbonneau \& MacGregor \cite{ch93}; Eggenberger et al. \cite{eg05b}) 
and internal gravity waves (Zahn et al. \cite{za97}; Talon \& Charbonnel \cite{ta05}). 
However, the question remains open as to how rotation should be modeled in a slightly more massive solar-type star
like \object{$\beta$ Virginis}. It is therefore worth investigating
how a flat rotation profile will
affect our conclusions about the effects of rotation on the small separations and
the rotational splittings of \object{$\beta$ Virginis}. 
The near solid body rotation of models including magnetics fields results in a 
low value for the diffusion coefficient associated to the shear turbulent mixing, 
and suggests that, for slow rotating solar-like stars, rotational mixing is less efficient in magnetic models
than in models with only rotation (Eggenberger et al. \cite{eg05b}). In the same way, the inclusion of internal gravity waves 
is found to reduce rotational mixing together with differential rotation (Talon \& Charbonnel \cite{ta05}).
The mixing in the central layers for a model of \object{$\beta$ Virginis} exhibiting a solar-like rotation profile will
thus be reduced resulting in a less pronounced increase in $\delta \nu_{02}$ and $r_{02}$ 
with the rotational velocity. As mentioned above, the values of the rotational splittings will also be influenced:
the rotational splittings computed for a model of \object{$\beta$ Virginis} with a uniform rotation velocity 
are smaller than the splittings 
corresponding to the rotation profile of our M1 model (mean value of 0.57\,$\mu$Hz instead of 0.83\,$\mu$Hz).
This results in a smaller scatter in the frequencies of non-radial modes for models
with a solar-like rotation profile than for our M1 model.
It is, however, clear that the accuracy of the present asteroseismic observations does not enable us to distinguish
between these models with different rotation profiles. 
We can simply hope that future asteroseismic observations will lead to 
accurate determination of rotational splittings and small separations for stars with various masses and ages,
in order to provide us with new insight into the rotation of solar-type stars 
and with constraints on how to model it best.

\section{Conclusion}
\label{conc}

The aim of this work was to determine the best model for the F9\,V star \object{$\beta$ Virginis} using the
Geneva evolution code.
By combining the existing non-asteroseismic observables with the new seismological measurements
by Carrier et al. (\cite{ca05b}), 
we find that two different solutions reproduce all these constraints well:
a main-sequence model with a mass of $1.28 \pm 0.03$\,$M_{\odot}$, an age $t=3.24 \pm 0.20$\,Gyr, and an initial metallicity
$(Z/X)_{\mathrm{i}}=0.0340 \pm 0.0040$, or else a model in the post-main  
sequence phase of evolution with a lower mass of $1.21 \pm 0.02$\,$M_{\odot}$ 
and a larger age $t=4.01 \pm 0.30$\,Gyr. Although these two models are compatible with all
observables, we note that the main-sequence model agrees slightly better with
the non-asteroseismic observables than the post-main-sequence one. 

We show that the small spacings $\delta \nu_{02}$ and the ratio $r_{02}$ between small and
large spacings are sensitive to the differences in the structure of the central layers between the main-sequence and
the post-main-sequence model; therefore, they can be used to unambiguously determine the evolutionary stage of \object{$\beta$ Vir}.
Unfortunately, existing asteroseismic data do not enable such a determination,
since none of the observed $\delta \nu_{02}$ and $r_{02}$ are available for 
frequencies where significant differences between the values of $\delta \nu_{02}$ and $r_{02}$ for main-sequence 
and post-main-sequence models are predicted.        

The effects of rotation were also studied.
We first find that the changes in the inner structure of the star due to the rotational mixing may be revealed by asteroseismic
observations, if precise measurements of $\delta \nu_{02}$ and $r_{02}$ can be made. 
We also show that the scatter of frequencies introduced by the rotational splittings can account for the larger dispersion
of the observed large spacings for the non-radial modes than for the radial modes

We conclude that the combination of non-asteroseismic measurements with existing frequency observations puts
important constraints on the global parameters of \object{$\beta$~Vir}, but that additional asteroseismic measurements are 
needed (especially measurements of the small spacings) to unambiguously determine its evolutionary stage
and to investigate the internal structure of this star in detail.

\begin{acknowledgements}
We thank J. Christensen-Dalsgaard for providing us with the Aarhus adiabatic pulsation code.
We also thank A. Maeder, G. Meynet, C. Charbonnel, and S. Talon for helpful advice. 
This work was supported financially by the Swiss National Science Foundation.
\end{acknowledgements}

\end{document}